\documentclass[aps,amsfonts,doublespace]{revtex4}
\usepackage{psfrag,graphicx,amstext,amsmath}
\usepackage{graphicx}
\usepackage[subnum]{cases}
\renewcommand{\huge}{\large}
\renewcommand{\LARGE}{\large}
\renewcommand{\Large}{\large}
\begin{document}

\def\be{\begin{equation}}
\def\ee{\end{equation}}
\def\bea{\begin{eqnarray}}
\def\eea{\end{eqnarray}}
\def\bml{\begin{mathletters}}
\def\eml{\end{mathletters}}
\def\l{\label}
\def\b{\bullet}
\def\eqn#1{(~\ref{eq:#1}~)}
\def\no{\nonumber}
\def\av#1{{\langle  #1 \rangle}}

%=============================================================================
%=============================================================================
\title{Extreme value distribution for weakly correlated fitnesses in
  block model}

\author{Kavita Jain}
\email{jain@jncasr.ac.in}
\affiliation{Theoretical Sciences Unit and Evolutionary and Organismal Biology Unit, Jawaharlal Nehru Centre for Advanced Scientific Research,  Jakkur P.O., Bangalore 560064,India}
\widetext
\date{\today}

\renewcommand{\huge}{\large}
\renewcommand{\LARGE}{\large}
\renewcommand{\Large}{\large}
\def\be{\begin{equation}}
\def\ee{\end{equation}}
\def\bea{\begin{eqnarray}}
\def\eea{\end{eqnarray}}
\def\no{\nonumber}
\thispagestyle{empty}

%=============================================================================
%=============================================================================
\begin{abstract}{We study the limit distribution of the largest
    fitness for weakly correlated 
and identically distributed random fitnesses. A fitness variable is
obtained by taking a linear combination of a fixed number of independent
  random variables drawn from the same parent distribution and two
  fitnesses are correlated if they have at least one common term in
  the respective sum. We find that for certain class of parent
  distributions, the extreme 
  value distribution for correlated random variables can be related
  either to 
  one of the known limit laws for independent 
  variables or the parent 
distribution itself. For other cases, new limiting distributions 
appear.  The conditions under which these results hold are
identified. 
}
\vskip0.5cm
\end{abstract}
\maketitle

%=============================================================================
\section{Introduction}
\label{intro}

Extreme value theory \cite{David:2003} deals with the
smallest or the largest of a set of random variables $\{x_1,...,
x_N\}$ and has found numerous applications in physics
\cite{Clusel:2008}, engineering \cite{Katz:2002}, 
biology \cite{Gillespie:1991} and finance \cite{Embrechts:2005}. 
If the $x_i$'s are independent and identically
distributed (i.i.d.) according to a continuous 
distribution $p(x)$, the probability of the $k$th maximum  is given by \cite{David:2003} 
\be
{\tilde P}^{(k)}_N(x)= \frac{(1-q(x))^{k-1} p(x) q(x)^{N-k}}{B(k,N-k+1)} 
\label{kmax}
\ee
where the cumulative distribution $q(x)=\int^x_0 dy
p(y)$ and $B(m,n)$ is the beta function. 
 A classic result in extreme value theory of i.i.d. random variables
 states that for large $N$ (and $k \ll N$), the distribution ${\tilde
   P}^{(k)}_N(x)$ is of the following scaling form \cite{David:2003}: 
\be
{\tilde P}^{(k)}_N(x) \approx \frac{1}{{\tilde a}_N} {\tilde
  F}^{(k)}\left(\frac{x-{\tilde b}_N}{{\tilde a}_N} \right)
\label{1max}
\ee
In the above expression, while the location factor ${\tilde b}_N$ and the scale
factor ${\tilde a}_N$ 
depend on the details of the parent distribution $p(x)$, the
scaling function ${\tilde
  F}^{(k)}$ is determined by the large $x$ behavior of
$p(x)$. For $k > 1$, the distribution  ${\tilde P}^{(k)}_N(x)$ can be 
related to ${\tilde P}^{(1)}_N$ by a transformation and the maximum
value distribution itself can be one of the following three types
\cite{David:2003}:

1. Fr{\'e}chet distribution if $p(x)$ decays as a power law

2. Weibull distribution if $p(x)$ is bounded above

3. Gumbel distribution if $p(x)$ is unbounded above and decays faster
   than a power law

Much  less is known about the extreme value statistics when the random
variables are not independent. It is interesting to investigate how
the correlations affect the limiting distributions of the three
i.i.d. classes mentioned above.  
If the correlations are strong, one may expect the extreme value
distribution to be different from the i.i.d. limit laws and universal 
distributions may not even exist 
\cite{Clusel:2008, Seetharaman:2010}. For weakly correlated variables,
on the other hand, the i.i.d. results may still hold.   
In this article, we study the extreme value statistics of random
variables with weak correlations and identify the relevant parameters
depending on which the extreme value distribution is seen to be 
either one of the suitably rescaled  
i.i.d. limit distributions or the (rescaled) parent
distribution or a completely 
different distribution unrelated to these.  

Extreme value 
statistics of correlated random 
variables has been studied  for stationary Gaussian process with
correlations decaying logarithmically or faster 
\cite{Leadbetter:1983} and specific physical examples such as random
energy model 
with correlated random potential \cite{Carpentier:2001,Fyodorov:2009},
$1/f$ noise \cite{Antal:2001}, directed
polymer on Cayley tree \cite{Dean:2001},
fluctuating interfaces \cite{Majumdar:2004} and mass transport models 
\cite{Evans:2008}. Here we study this 
problem in the context of biological evolution where the 
question of extreme values naturally arises as a large
population evolves to maximise its fitness. During the first step in
evolution, an individual in an initially unfit population may
acquire one or more mutations. If the total 
population is infinitely large, the subpopulation with $D$ mutations
is dominated by the {\it fittest} mutant in this 
subpopulation \cite{Jain:2005}. 
As several experiments have shown 
that the fitnesses are not completely random
\cite{Perelson:1995,Carneiro:2010,Miller:2011}, we 
are led to study the extreme value statistics of correlated
fitnesses.

In order to examine the implications of correlations amongst
fitnesses, several models 
such as $NK$ model \cite{Kauffman:1993}, block model 
\cite{Perelson:1995} and rough Mt. Fuji-type model \cite{Aita:2000}
have been proposed. In this article, we will work with the block model
which has been employed in recent
years to address questions pertaining to 
evolutionary dynamics 
\cite{Orr:2006,Welch:2005,Kryazhimskiy:2009}. The extreme value
statistics in the block model has been studied for both strongly and weakly
correlated fitnesses for an exponentially decaying parent distribution $p(x)$
\cite{Jain:2009}.  
However the question of the universality of the extreme value
distributions was not addressed. Here we extend the
previous study by considering a broad class of parent distributions
and focus on weakly 
correlated but identically distributed fitnesses.

The rest of the article is organised as follows. The block model of
correlated fitnesses is defined in  
Sec.~\ref{model}. The extreme value distributions are obtained when a
single mutation is present in the initial fitness in Sec.~\ref{D1} and
in the presence of multiple mutations  in Sec.~\ref{Dm}. Finally a
discussion and summary is given in Sec.~\ref{con}.

%=============================================================================

\section{Block model of correlated fitnesses}
\label{model}

\begin{figure}[t]
\begin{center}
\includegraphics[angle=0,scale=0.7]{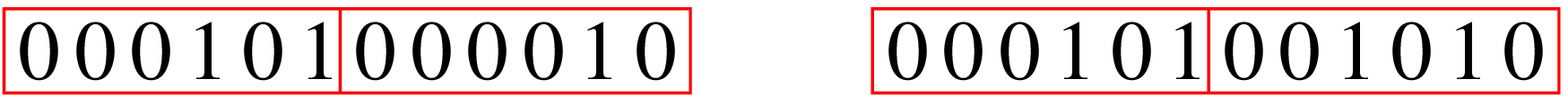}
\caption{Block model: A binary sequence of length $L=12$ is divided
  into two blocks of equal length $\ell=6$. The sequences on the left
  and right differing by single mutation have correlated fitnesses as 
  the first block is common between the two sequences.}
\label{fmodel}
\end{center}
\end{figure}

Many biomolecules such as proteins, antibodies and enzymes have 
natural domains or partitions and can be modeled as a sequence of
length $L$ divided into several blocks  
\cite{Perelson:1995}.    
In the simplest setting, a sequence represented by a binary
string of zeros and ones is divided into $B$ blocks of equal length 
$\ell=L/B~,~1 \leq \ell \leq L$ (see Fig.~\ref{fmodel}). A block
configuration is assigned a  
random  fitness value regardless of its position in the sequence. 
These block fitnesses are
chosen {\it independently} from a common distribution $p(f)$ which is
nonzero on the interval $[a,b]$ and zero elsewhere. Then the sequence
fitness  is obtained by averaging over the fitnesses of the blocks in
the sequence. 

If two sequences have several blocks in common, their fitnesses will
also be similar and hence correlated. 
An attractive feature of the block model is that the fitness 
correlations can be tuned with the block length $\ell$. For $\ell=1$,
as two distinct sequences can have up to $L-1$ blocks in common,
sequence fitnesses are maximally correlated while for  
$\ell=L$, we obtain the model with maximally uncorrelated fitnesses as
no common blocks are possible \cite{Jain:2005}. 
To see how fitness correlations vary with block length, consider a set of 
sequences carrying one mutation relative to the sequence with
all zeros. Then as a single
mutation in this sequence leaves $B-1$ 
blocks unchanged, the 
fitness $w_j$ of the one mutant neighbor of the initial sequence is given by 
\be
w_j=\frac{(B-1) f_0+f_{j}}{B} ~,~j=1,...,\ell \label{d1defn}
\ee
where $f_0$ and $f_j$ denote the fitness of the block configuration
with all zeros and with $\ell-1$ zeros and a one at the $j$th locus
respectively. Using the fact that the block fitnesses are
independently distributed, we find that the correlation amongst the
sequence fitnesses $\{w_j \}$ is given by 
\bea
\langle w_i w_j \rangle - \langle w_i \rangle \langle w_j \rangle =
\left[\frac{(B-1)^2+\delta_{i,j}}{B^2} \right]\sigma^2=
\left[ \left(1-\frac{\ell}{L}\right)^2 +\delta_{i,j}
  \left(\frac{\ell}{L} \right)^2 \right]\sigma^2
\eea
where $\sigma^2$ is the variance of the block fitness
distribution. It is clear from the last equation that the correlations 
decay as the block length $\ell$ increases towards $L$. In this article, we
are interested in weakly correlated fitnesses which are obtained when $L 
\to \infty, \ell \to \infty$ with $B$ fixed.

%=============================================================================
\section{Distribution of the largest fitness}

Starting from an initial unfit sequence composed of 
identical blocks (say, all zeros), we are interested 
in finding the distribution of the largest fitness amongst ${L \choose
  D }$ sequences with $D$ mutations
relative to the initial sequence. In this article, we focus on the 
extreme value distribution for  {\it nonindependent and identically
  distributed} (ni.i.d.) random  
variables (for some results on nonindependent and nonidentically 
distributed fitnesses, see \cite{Jain:2009,Seetharaman:2010}). 
Such ni.i.d. sequence fitnesses 
  are obtained when either the number of mutations $D=1$ for any
  $B > 1$ or the number of blocks $B=2$ and $D$ is odd. 
In the first case (discussed in Sec.~\ref{D1}), the distribution of
sequence fitness $w_j$ defined by (\ref{d1defn}) is given by
\be
\textrm{Prob}(w_j)= \int_0^\infty \int_0^\infty df_0 df_j p(f_0)
p(f_j) \delta \left(w_j- \frac{(B-1) f_0+f_{j}}{B}\right)
\label{seqfit}
\ee
which is independent of $j$ and thus the sequence fitnesses with
single mutation are identically
distributed. In the second case when $B=2$ and a sequence has $D$
mutations, the sequence fitness is 
obtained by averaging over the fitness of the first block with 
$d$ mutations and the second block with $d'=D-d$ mutations (see
Sec.~\ref{Dm}). On writing down the 
sequence fitness distribution similar to (\ref{seqfit}), it is readily 
verified that only those sequences with different block configuration
in first and 
second block have identically distributed fitness. Since a sequence
with same configuration in both blocks can occur only for even $D$, it
follows that ni.i.d. fitnesses are obtained when $D$ is odd.  
In the rest of the article, we assume $L$ to be an  
even integer and consider $D \leq L/2$ as the results for $D > L/2$
can be obtained on replacing $D$ by $L-D$. 

%--------------------------------------------------------------------------

\subsection{Single mutation in the initial sequence}
\label{D1}

\begin{figure}[t]
\begin{center}
\begin{psfrags}
\psfrag{x}[B][B][2.5]{$(2 w-{\tilde b}_\ell)/{\tilde a}_\ell$}
\psfrag{y}[L][L][2.5]{${\tilde a}_\ell P_\ell^{(1)}(w)/2$}
\includegraphics[angle=270,scale=0.34]{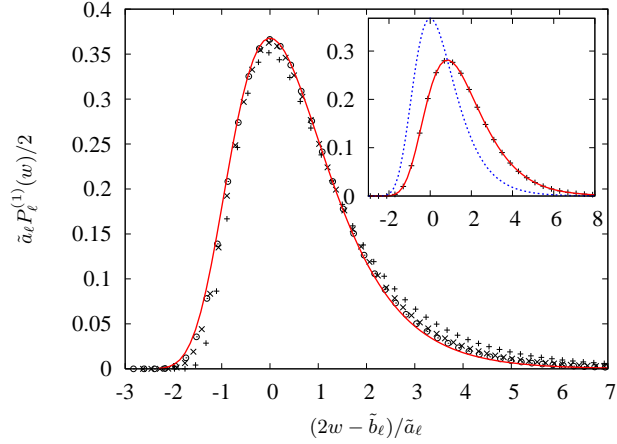}
\end{psfrags}
\caption{Scaled distribution for the first maximum when the block
 fitness distribution $p(f)=e^{-\sqrt{f}}/(2 \sqrt{f})$ 
 (main) and $e^{-f}$ (inset) for $B=2$. The points
  refer to exact integration of 
  (\ref{D1kmax}) for  $\ell=10^{2} (+),
  10^{5} (\times)$ and $10^{10} (\circ)$. The solid lines show 
 (\ref{D1univ}) in main and (\ref{D1exact}) in inset. 
The inset also shows the Gumbel distribution for comparison (dotted curve).}
\label{D1univexpo}
\end{center}
\end{figure}

We first consider the extreme value statistics of the fitness set
$\{w_j\}$ defined by (\ref{d1defn}).  
For a given $f_0$, the 
sequence fitness $v= ((B-1) f_0+ f)/B$ is the largest
amongst the set $\{w_j \}$ 
if $f=\max\{f_1,...,f_\ell \}$. But as the block fitnesses 
are i.i.d. random variables, this event occurs with a probability
${\tilde P}^{(1)}_\ell(f)$. Then the probability $P^{(1)}_\ell(w)$
that the largest fitness in the set $\{ w_j \}$  has a value $w$ can
be written as   
\be
P^{(1)}_\ell(w)=\int_0^{\infty}\int_0^{\infty} df_0 df p(f_0) {\tilde
  P}^{(1)}_\ell(f)
\delta(w-v) 
\ee
In general, the probability $P^{(k)}_\ell(w)$ that the $k$th
maximum has a value $w$ is given by
\bea
P^{(k)}_\ell(w) 
&=& \int_0^{\infty}\int_0^{\infty} df_0 df p(f_0) {\tilde
  P}^{(k)}_\ell(f)
\delta(w-v) \no \\
&=& \frac{B}{B-1} \int_{-\infty}^{\infty} df p\left(\frac{B w-f}{B-1}
\right){\tilde P}^{(k)}_\ell(f) 
\label{D1kmax}
\eea
For large $\ell$, using (\ref{1max}) we obtain
\be
P^{(k)}_\ell (w) 
\approx  \frac{B}{B-1} \int_{-\infty}^{\infty} df p\left(\frac{B w-f}{B-1}
\right) ~\frac{1}{{\tilde a}_\ell}{\tilde F}^{(k)}\left(\frac{f-{\tilde b}_\ell}{{\tilde a}_\ell}
\right)
\label{D1kmaxL}
\ee
The behavior of the distribution function in the above equation can be
classified as follows: 

(i) If ${\tilde a}_\ell$ diverges with $\ell$, it is useful to
rewrite  (\ref{D1kmaxL}) as 
\bea
P^{(k)}_\ell(w) \approx B \int_{-\infty}^\infty dz p(z)
~\frac{1}{{\tilde a}_\ell}{\tilde F}^{(k)}\left(\frac{(1-B) z+ B w-{\tilde b}_\ell}{{\tilde a}_\ell}
\right) 
\eea
In the limit $\ell, w \to \infty$, the ratio $z/{\tilde a}_\ell$ in the
argument of ${\tilde F}^{(k)}$ above can be ignored and we can write  
\be
P^{(k)}_\ell(w) \approx   \frac{B}{{\tilde a}_\ell} {\tilde   F}^{(k)} \left(\frac{B
  w-{\tilde b}_\ell}{{\tilde a}_\ell}\right)
\label{D1univ}
\ee
where we have used that $p(z)$ is normalised to unity. Thus up to a
rescaling, the extreme value distribution function for 
 correlated variables falls in the same
 universality class as the i.i.d. ones if the scale factor ${\tilde
   a}_\ell$ diverges. The scaling variable in (\ref{D1univ}) indicates
 that the distribution decays faster for correlated variables ($B >
 1$) as one would 
 intuitively expect. An example of a class of block fitness
 distribution for which ${\tilde a}_\ell$ diverges with $\ell$ is 
 $p(f)= \delta f^{\delta-1} 
 e^{-f^\delta}$. In this case, the Gumbel scaling function ${\tilde
 F}^{(1)}(y)=e^{-y} e^{-e^{-y}}$ with the location factor ${\tilde
   b}_\ell=(\ln \ell)^{\frac{1}{\delta}}$ and 
 the scale factor 
 ${\tilde a}_\ell= \delta^{-1} (\ln \ell)^{\frac{1-\delta}{\delta}}$ which
 diverges with $\ell$ for $0 < \delta < 1$ \cite{David:2003}. A good
 agreement is seen between the 
 exact distribution (\ref{D1kmax}) and asymptotic result 
 (\ref{D1univ}) in Fig.~\ref{D1univexpo} for $\delta=1/2$.

\begin{figure}[t]
\begin{center}
\begin{psfrags}
\psfrag{x}[B][B][2.5]{$2 w-{\tilde b}_\ell$}
\psfrag{y}[L][L][2.5]{$P_\ell^{(1)}(w)/2$}
\includegraphics[angle=270,scale=0.34]{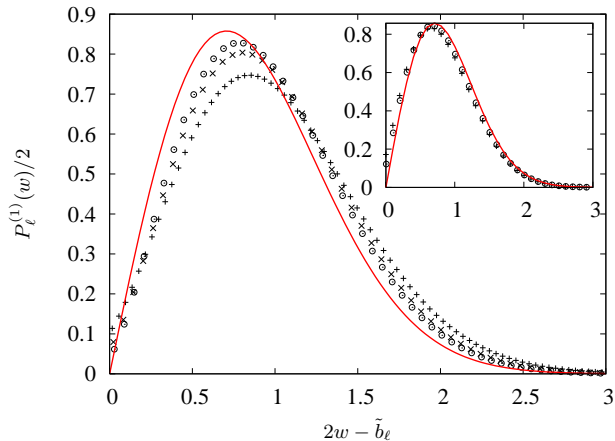}
\end{psfrags}
\caption{Scaled distribution for the first maximum (main) and the second
  maximum (inset) for the block fitness distribution $p(f)=2 f
  e^{-f^2}$ and $B=2$. The plot shown in points
  refer to exact integration of (\ref{D1kmax}) for  $\ell=10^{2} (+),
  10^{5} (\times)$ and $10^{10} (\circ)$ and the 
  lines to (\ref{D1non}).}
\label{D1nonuniv}
\end{center}
\end{figure}

(ii) If the scale factor ${\tilde a}_\ell$ vanishes as $\ell \to
\infty$, a change of variables in (\ref{D1kmaxL}) for $B > 1$ gives 
\bea
P^{(k)}_\ell(w) 
&\approx&  \frac{B}{B-1} \int_{-\infty}^\infty dz {\tilde F}(z)
p\left( \frac{{\tilde 
    a}_\ell z+{\tilde b}_\ell-B w}{1-B} \right) \\
&\approx& \frac{B}{B-1} p\left(\frac{B w-{\tilde
    b}_\ell}{B-1}\right)~,~B \neq 1
\label{D1non}
\eea
where the last expression is obtained in the limit $\ell, w \to \infty$. 
 Thus the probability distribution of the $k$th maximum is a rescaled
 parent distribution  and independent of $k$ if ${\tilde a}_\ell$ goes
 to zero with increasing $\ell$. 
 Figure \ref{D1nonuniv} shows that for Gaussian distributed block
 fitnesses, the asymptotic 
 distribution in (\ref{D1non}) approaches the exact distribution in
 (\ref{D1kmax}) as $\ell$ increases. 

(iii) 
If the scale factor ${\tilde a}_\ell$ is 
independent of $\ell$ for large $\ell$, one may expect new limiting
distributions to arise.  
For example, for exponentially distributed block fitnesses 
($\delta=1$), the scale factor is unity and the limiting distribution
of the $k$th maximum for i.i.d. variables is given by \cite{David:2003}
\bea
 {\tilde
  P}^{(k)}_\ell(x) \approx {\tilde
  F}^{(k)}(y)=\frac{\textrm{exp}\left(-e^{-y}-k y \right)}{(k-1)!}
 \sim \left\{
\begin{array}{ll}
e^{-ky} ~,~ y \to \infty  \\
e^{-e^{-y}}~,~ y \to -\infty 
\end{array}
\right.
\label{expok}
\eea
where $y=x-\ln \ell$. 
Using this result in (\ref{D1kmaxL}) for  
large $\ell$, we can write
\bea
P^{(k)}_\ell(w) 
&\approx & \frac{B}{B-1} \int_0^{B w} df e^{-\frac{B w-f}{B-1}} e^{-\ell
  e^{-f}} \frac{(\ell e^{-f})^k}{(k-1)!} \no \\
&=&  \frac{B}{B-1} \frac{(\ell e^{-B
  w})^{\frac{1}{B-1}}} {(k-1)!}\int_{\ell e^{-B
  w}}^\ell df e^{-f} f^{k-\frac{B}{B-1}}  \no \\
&\approx& \frac{B}{B-1} \frac{e^{\frac{-u}{B-1}}}{(k-1)!} \Gamma \left(k-
\frac{1}{B-1},e^{-u} \right)
\label{D1exact}
\eea
where the last expression is obtained in the scaling limit $\ell,w \to
\infty$ keeping $u=B w  -\ln \ell$ finite and $\Gamma(a,z)$ is the 
  incomplete gamma function 
\cite{Gradshteyn:1980}. 
When $u \to -\infty$, an asymptotic expansion for large $e^{-u}$
  shows that $P_\ell^{(k)}(w) \sim e^{-e^{-u}}$ and thus the
  behavior of the backward tail is similar to that in the
  i.i.d. case. On the other hand, when $u \to \infty$, the
  incomplete gamma function is well approximated by the complete gamma
  function $\Gamma \left(k-
\frac{1}{B-1},0 \right)$ except for $k=1$ and $B=2$ and thus gives
  $P_\ell^{(k)}(w) \sim e^{-u/(B-1)}$. For $k=1$ and $B=2$, the
  incomplete gamma function $\Gamma(0, 
  e^{-u}) \approx u-C$ where $C$ is Euler-Mascheroni
  constant but the tail of the maximum value distribution decays
  exponentially fast in this case also. A comparison of the 
  distribution $P_\ell^{(1)}(w)$ with the Gumbel distribution is shown
   in the inset of Fig.~\ref{D1univexpo}.

%---------------------------------------------------------------------------
\subsection{Multiple mutations in the initial sequence}
\label{Dm}

We now turn to the case when a sequence is composed of two blocks and
carries odd $D \geq 1$ mutations. For a sequence divided into two blocks of
equal length, the sequence fitness is 
obtained by averaging over the fitness of the first block with 
$d$ mutations and the second block with $d'=D-d$ mutations. Although
the integer $d$ runs from $0$ to $D$, 
distinct sequence fitnesses are obtained if $d \leq d_u=(D-1)/2$. As
the first block can have $s_d={\ell \choose d}$ independent fitnesses
and the second block can have $s_{d'}$ fitnesses, the number of 
distinct sequence fitnesses is given by $ 
  N_D= \sum_{d=0}^{d_u} s_d s_{d'}=(1/2) {L \choose D}$
  \cite{Gradshteyn:1980} which reduces to $\ell$ for $D=1$.

We are interested in the distribution $P^{(1)}_{N_D}(w)$ of the
largest fitness $w$ among such $N_D$ correlated fitnesses.  It is
convenient to work with the cumulative distribution ${\cal
  P}_{N_D}^{(1)}(w)$ defined as 
\be
{\cal
  P}_{N_D}^{(1)}(w)=\int_0^w dw' P^{(1)}_{N_D}(w')
\ee
which gives the probability that all the  $N_D$ fitnesses are smaller
than $w$. If 
 ${\cal G}_{s_d}(w)$ denote the probability that the fitness of all
the sequences carrying $d$ mutations in the first block and $d'$ in the
second block is smaller than $w$, we can write
\be
{\cal
  P}_{N_D}^{(1)}(w)=\prod_{d=0}^{d_u} {\cal G}_{s_d} (w)
\label{mult}
\ee
Our task thus reduces to finding the distribution ${\cal G}_{s_d}
(w)$ which can be calculated in a manner similar to the single
mutation case. For a given $d$, the largest sequence 
fitness $v=(f_d+f_{d'})/2$ is obtained 
when both the block fitnesses $f_d$ and $f_{d'}$ of the first and
second block respectively are the largest amongst the set of
$s_{d}$ and $s_{d'}$ i.i.d. random variables. Since the probability
${\cal G}_{s_d}(w)$ equals the probability that $v < w$, we get 
\bea
{\cal G}_{s_d}(w) 
&=& \int_0^\infty df_d \int_0^\infty df_{d'} {\tilde
  P}^{(1)}_{s_{d}}(f_{d}) {\tilde
  P}^{(1)}_{s_{d'}}(f_{d'}) \Theta(w-v) \nonumber \\
&=& \int_0^{\infty} df_{d}{\tilde
  P}^{(1)}_{s_{d}}(f_{d}) {\tilde {\cal P}}^{(1)}_{s_{d'}}(2 w-f_{d}) \label{G1}\\
&=& \int_0^{\infty} df_{d'}{\tilde {\cal P}}^{(1)}_{s_{d}}(2
  w-f_{d'}) {\tilde
  P}^{(1)}_{s_{d'}}(f_{d'}) 
\label{G2}
\eea
where the distribution $\tilde {\cal P}(w)=\int_0^w dw'  \tilde {P}(w')$. 

\begin{figure}[t]
\begin{center}
\begin{psfrags}
\psfrag{x}[B][B][2.5]{$2 w/{s_D}$}
\psfrag{y}[L][L][2.5]{$s_D P^{(1)}(w)/2$}
\includegraphics[angle=270,scale=0.34]{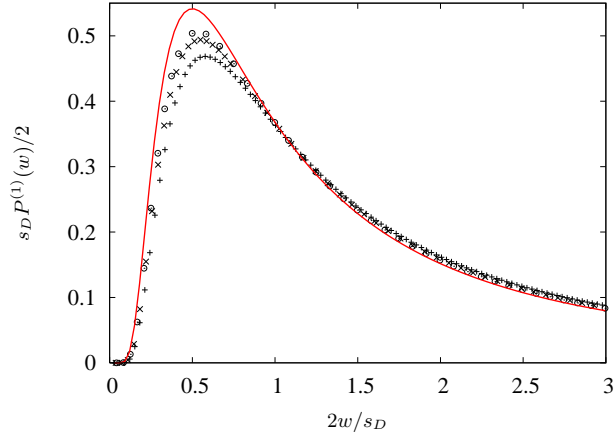}
\end{psfrags}
\caption{Scaled distribution for the first maximum for $p(f)=f^{-2}
  e^{-1/f}$ and $D=5$ obtained using exact integration of (\ref{mult})
  for  $\ell=40 (+), 
  60 (\times)$ and $80 (\circ)$. The solid line shows the Fr{\'e}chet
  scaling function ${\tilde F}^{(1)}(y)=y^{-2} e^{-1/y}$.} 
\label{ellhalfpwr}
\end{center}
\end{figure}

In the limit $\ell \to \infty$, the distributions ${\cal G}_{s_0}(w)$
and ${\cal G}_{s_d}(w), d > 0$ need to be analysed separately. 
In the former case, since $s_0=1$ on using (\ref{kmax}) and
(\ref{1max}) in (\ref{G2}) we have 
\bea
{\cal G}_{s_0}(w) \approx \int_{-\infty}^\infty df q(2 w-f)
\frac{1}{{\tilde a}_{s_D}}{\tilde F}^{(1)}\left(\frac{f-{\tilde
    b}_{s_D}}{{\tilde a}_{s_D}}\right) 
\label{Gs0}
\eea
which is simply the cumulative maximum value distribution for 
$s_D$ random variables when there is a single
mutation in the initial sequence and therefore exhibits a behavior
similar to that already discussed in Sec.~\ref{D1}. 
For $d > 0$, as both $s_d, s_{d'}$ are 
large for $\ell \gg 1$, using (\ref{1max}) in (\ref{G1}) we get 
\bea
{\cal G}_{s_d}(w) &\approx&  \int_{-\infty}^{\infty} df \frac{1}{{\tilde
    a}_{s_d}} {\tilde
  F}^{(1)} \left(\frac{f-{\tilde b}_{s_d}}{{\tilde a}_{s_d}}\right)
{\tilde
  {\cal F}}^{(1)} \left(\frac{2 w -f-{\tilde b}_{s_{d'}}}{{\tilde
    a}_{s_{d'}}}\right) \\
&\approx & {\tilde {\cal F}}^{(1)} \left(\frac{2 w-{\tilde b}_{s_{d}}-{\tilde
b}_{s_{d'}}}{{\tilde a}_{s_{d'}}} \right) 
\eea
where ${\tilde {\cal F}}(w)= \int_0^w dw' {\tilde {F}}(w')$ and the
last expression is obtained if ${\tilde a}_{s_d}/{\tilde a}_{s_{d'}}
\to 0$ as $\ell \to \infty$. The distribution  ${\cal G}_{s_d}(w)$ can
be found in a similar manner in the opposite limit ${\tilde 
  a}_{s_{d'}}/{\tilde a}_{s_{d}} \to 0$ and we have 
\begin{numcases}
{{\cal G}_{s_d}(w) \approx }
%\left\{\begin{array}{ll}
{\tilde {\cal F}}^{(1)} \left(\frac{2 w-{\tilde b}_{s_{d}}-{\tilde
b}_{s_{d'}}}{{\tilde a}_{s_{d'}}} \right)~,~
\text{if~} {\tilde a}_{s_d}/{\tilde a}_{s_{d'}} \to 0  \label{Gell1}\\
 {\tilde {\cal F}}^{(1)}\left(\frac{2 w-{\tilde b}_{s_{d}}-{\tilde
b}_{s_{d'}}}{{\tilde a}_{s_{d}}} \right) ~,~
\text{if~} {\tilde a}_{s_{d'}}/{\tilde a}_{s_{d}} \to 0
\label{Gell2}
\end{numcases}
Thus if the appropriate scale factor ratio vanishes for all $d \leq
  d_u$, one can express ${\cal 
  P}_{N_D}^{(1)}(w)$ as a product of the parent distribution $q$
  and the i.i.d. functions ${\tilde 
  {\cal F}}$. Otherwise a novel distribution may be expected. If each
  integral in the product on the right hand side 
  (RHS) of
  (\ref{mult}) can be calculated,  
a further step is required for $D > 1$ to ascertain if there is a
single scaling 
variable and the product  is  reducible to a single
function.  To illustrate these points, we now apply the above discussion
to specific block fitness distributions: 

(i) For a class of block fitness distributions $p(f)= \gamma
e^{-f^{-\gamma}} f^{-1-\gamma}$, $\gamma > 0$ which decay as a power
law, the location factor 
${\tilde b}_N=0$ and the scale factor ${\tilde a}_N=N^{1/\gamma}$ so that the 
ratio ${\tilde a}_{s_d}/{\tilde 
  a}_{s_{d'}}=(s_{d}/s_{d'})^{1/\gamma}$. As described below, two
distinct cases arise  depending on whether the number of mutations $D$
is comparable to $\ell$:

Case a.  When $D/\ell \to 0$
as $\ell \to \infty$, using the approximation $s_d \approx \ell^d/d!$,
we find that  ${\tilde a}_{s_d}/{\tilde 
  a}_{s_{d'}} \sim \ell^{2d-D}$ which vanishes for all $d \leq d_u$ for
large $\ell$.  Then due to (\ref{Gell1}), the cumulative distribution ${\cal
  P}_{N_D}^{(1)} (w)$ for large $\ell$ can be written as 
\be
{\cal
  P}_{N_D}^{(1)} (w)\approx \prod_{d=0}^{d_u} {\tilde {\cal
    F}}^{(1)}\left( \frac{2 w}{{\tilde a}_{s_{d'}}}\right)=
\prod_{d=d_u+1}^D  {\tilde {\cal F}}^{(1)}\left( \frac{2 w}{{\tilde a}_{s_d}}\right)
\ee
where the cumulative Fr{\'e}chet distribution ${\tilde {\cal
    F}}^{(1)}(y)=e^{-y^{-\gamma}}~,y > 0$ and zero otherwise \cite{David:2003}.
In the limit $w, \ell \to \infty$ with $W_n= 2 w/{\tilde
  a}_{s_{n}}$ fixed, as the Fr{\'e}chet scaling function
\begin{numcases}
{{\tilde {\cal F}}^{(1)} \left( \frac{2 w}{{\tilde
  a}_{s_{m}}} \right) ={\tilde {\cal F}}^{(1)} \left( \frac{W_n {\tilde
  a}_{s_{n}}}{{\tilde
  a}_{s_{m}}} \right) \to } 1 \textrm{~if~} m < n \\
0 \textrm{~if~} m > n
\end{numcases}
a nontrivial extreme value distribution is obtained if $W_D=2
  w/{\tilde a}_{s_{D}}$ 
  is kept fixed. 
Thus we find that the maximum value distribution is given by 
\be
P_{N_D}^{(1)}(w) \approx \frac{2}{{\tilde
  a}_{s_{D}}} {\tilde F}^{(1)} \left( \frac{2 w}{{\tilde
  a}_{s_{D}}} \right) 
\ee
which is verified in Fig.~\ref{ellhalfpwr} for $D=5$ and
$\gamma=1$. As the above result holds for $D=1$ as well (see
  (\ref{D1univ})), we conclude that  
the universality class does not change from the i.i.d. class for
$1 \leq  D \ll \ell$ for parent distributions decaying as a power
  law. 

Case b.  For
finite $D/\ell$ in the large $\ell$ limit, we can write 
\begin{numcases}
{\frac{s_d}{s_{d'}} = \prod_{n=1}^{D-2 d} \frac{d+n}{\ell-D+d+n}
  \approx} \textrm{exp} \left[ \ell \int_0^{R-2 r} dx 
\ln \left(\frac{r+x}{1-R+r+x} \right) \right] ~,~  R-2 r > 0 \label{e1} \\ 
\left(\frac{r}{1-r} \right)^{D-2d}~,~  R-2 r =0 \label{e2}
\label{ratio}
\end{numcases}
where $r=d/\ell$ and $R=D/\ell$. Since the integrand in the
exponential on the RHS of
(\ref{e1}) is always negative, $s_d/s_{d'}$ decays to 
zero as $\ell$ increases for nonzero $R-2 r$ but remains finite for vanishing
$R-2 r$.  As a result, the integral in
(\ref{G1}) or (\ref{G2}) does not reduce to  i.i.d. distributions for
$d \sim d_u$ and
we may expect novel extreme value distributions for $D \sim \ell$.

(ii) A similar analysis can be carried out for 
  bounded distributions $p(f)= 
  \nu (1-f)^{\nu-1}~,~\nu > 0 $ for which ${\tilde b}_N=1$ and 
${\tilde a}_N=N^{-1/\nu}$. When the number of mutations $D=1$, a 
  nonuniversal parent distribution is obtained by replacing $\ell$ by
  $s_D$ in (\ref{D1non}) 
  while a rescaled universal  
  distribution holds for $1 < D \ll \ell$. Since the ratio ${\tilde
  a}_{s_{d'}}/{\tilde a}_{s_{d}} \to 0$ as $\ell \to \infty$,
  due to (\ref{D1non}) and (\ref{Gell2}), we get 
  \be
{\cal P}_{N_D}^{(1)} 
(w) \approx q(2 w-1) \prod_{d=1}^{d_u} {\cal {\tilde F}}^{(1)}\left( \frac{2 (w-1)}{{\tilde
  a}_{s_{d}}}\right)   
\ee
where  the Weibull distribution 
  ${\cal {\tilde F}}^{(1)}(y)=e^{-(-y)^\nu}, y \leq 0$ and unity
  otherwise \cite{David:2003}. As $w \to 1$ and $\ell \to \infty$,
  while the
  cumulative distribution $q \to 1$, the distribution 
${\cal {\tilde F}}^{(1)}\left[{2
  (w-1)}/{{\tilde a}_{s_{m}}}\right]$ approaches zero for $m > n$ and
  unity for $m < n$ if $W_n=2 (w-1)/{\tilde a}_{s_n}$ is
  kept fixed. Then a nontrivial maximum value distribution is obtained
  if $2 (w-1)/{\tilde a}_{s_{d_u}}$ remains finite and we obtain
\be
P_{N_D}^{(1)} 
(w) \approx \frac{2}{{\tilde a}_{s_{d_u}}} {\tilde F}^{(1)}\left( \frac{2 (w-1)}{{\tilde a}_{s_{d_u}}}
\right)  ~,~ 1 < D \ll \ell
\ee
However for $D \sim \ell$, new 
  extreme value distributions are expected for the reasons mentioned above.

(iii) We finally consider block fitness distributions for which the
  scale factor ratio  
  typically tends to a finite limit as $\ell \to \infty$. For $p(f)=
  \delta f^{\delta-1}  
  e^{-f^\delta}$ considered in Sec.~\ref{D1}, the 
  scale factor ratio ${\tilde a}_{s_{d}}/{\tilde 
  a}_{s_{d'}}=(\ln s_{d}/\ln s_{d'})^{(1-\delta)/\delta}~,~\delta \neq
  1$. Using
  Stirling's approximation for $d > 0$, we may write
\begin{numcases}
{\frac{\ln s_{d}}{\ln s_{d'}} \approx \frac{r \ln( (1-r)/r)-\ln (1-r)
  }{r' \ln( (1-r')/r')-\ln (1-r') } \to } \textrm{constant}~,~\textrm{if}~
  D/\ell \to 0 \\
0 ~,~\textrm{if} ~d/\ell \to 0, D/\ell ~\textrm{finite} \\
\textrm{constant}~,~\textrm{if both} ~d/\ell, D/\ell ~\textrm{finite} 
\end{numcases}
where $r=d/\ell$ and $r'=d'/\ell$. Thus we cannot express ${\cal P}^{(1)}(w)$
  in terms of the known functions even for $1 < D \ll \ell$ when
  $\delta \neq 1$. But some progress is possible for $\delta=1$
  which we discuss next. 

As we have already mentioned, for $p(f)=e^{-f}$ the scale factor is
independent of $\ell$ and thus the scale ratio is 
a constant for any $D$. However it is possible to find ${\cal
  G}_{s_d}$ for all $d \leq d_u, D \geq 1$ for exponentially
distributed fitnesses. Although the maximum value
distribution for this case has been studied in a previous work 
\cite{Jain:2009}, here we present a simpler derivation. 
 For large $\ell$, using
  (\ref{expok}) for $k=1$ in (\ref{G2}), we obtain
\bea
{\cal
  P}_{N_D}^{(1)}(w) &\approx& 
  \int_{0}^{2 w} df_0 (1-e^{-2 w+f_0}) s_D e^{-f_0} e^{-{s_D} e^{-f_0}}
 \prod_{d=1}^{d_u} \int_{0}^{2 w} df_d
e^{-s_{d} e^{-2 w +f_d}}  s_{d'} e^{- f_d} e^{-s_{d'} e^{-f_d}} 
\label{PND} \\
&=& \int_{s_D e^{-2 w}}^{s_D} df_0 e^{-{f_0}} \left(1- \frac{s_D e^{-2
    w}}{f_0} \right) 
\prod_{d=1}^{d_u} \int_{s_{d'} e^{-2 w}}^{s_{d'}} df_d e^{-f_d}
e^{-s_{d} s_{d'} e^{-2 w}/f_d} 
\eea
where we have used (\ref{Gs0}) for $d=0$ term. In the
limits $w, \ell \to \infty$, the above integrals are nontrivial 
if $W_{d}=s_{d} s_{d'} e^{-2 w}$ is finite and we get
\bea
{\cal
  P}_{N_D}^{(1)}(w) &\approx& \int_{W_0}^{\infty} df_0 e^{-{f_0}} \left(1- \frac{W_0}{f_0} \right) 
\prod_{d=1}^{d_u} \int_{0}^{\infty} df_d e^{-f_d}
e^{-W_d/f_d} \\
&=& \left[e^{-W_0} -W_0 \Gamma(0,W_0) \right] \prod_{d=1}^{d_u} 2 \sqrt{W_d} K_1 (2 \sqrt{W_d})
\label{err}
\eea
where $K_1(x)$ is the modified Bessel function of second kind \cite{Gradshteyn:1980}. One can
check that the above equation reduces to the cumulative distribution 
for $D=1$ obtained in Sec.~\ref{D1} on using $W_0=e^{-u}$ in
(\ref{D1exact}) \cite{note}. We now consider the above cumulative
distribution when $D/\ell$ is 
zero and when it is finite in the large $\ell$ limit.

\begin{figure}[t]
\begin{center}
\begin{psfrags}
\psfrag{x}[B][B][2.5]{$w$}
\psfrag{y}[L][L][2.5]{$$}
\includegraphics[angle=270,scale=0.34]{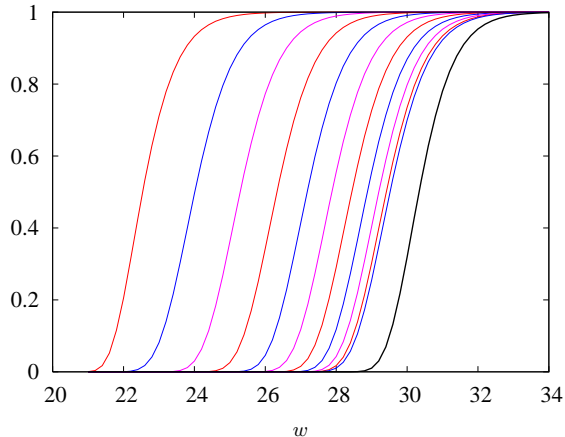}
\end{psfrags}
\caption{Cumulative distribution ${\cal P}^{(1)}(w)$ (bold line) for
  exponential parent distribution when $D=21$ and $\ell=80$ obtained using
  (\ref{PND}). The other curves show the individual factors ${\cal
  G}_{s_d}$ in the
  product in (\ref{PND}) with $d=0,...,d_u$ (left to right).} 
\label{expoDlarge}
\end{center}
\end{figure}

Case a. Since the distribution ${\cal G}_{s_d}$ moves towards larger
$w$ as $d$ increases (see Fig.~\ref{expoDlarge}), we may try $W_{d_u}$
as a scaling 
variable to reduce the product in (\ref{mult}) to a single
function. For $D \ll \ell$ on using $s_d \approx \ell^d/d!$ for large
$\ell$, we see that 
\be
\frac{W_d}{W_{d_u}} \approx \frac{d_u ! (D-d_u)!}{d ! (D-d)!}
\ee 
which is finite for all $d < d_u$. Thus unlike for the algebraically
decaying or bounded block fitness distributions, all the factors in (\ref{err})
contribute to the cumulative distribution ${\cal P}^{(1)}_{N_D}(w)$.

Case b. For $D \sim \ell$,  the ratio $W_d/W_{d_u}$ goes to
zero for nonzero $R-2 r$ and unity for vanishing $R-2r$:
\begin{numcases}
{\frac{W_d}{W_{d_u}} = \prod_{n=1}^{d_u-d}
\frac{d+n}{D-d_u+n} \times \frac{\ell-D+d+n}{\ell-d_u+n}  
  \approx} \textrm{exp} \left[\ell \int_0^{\frac{R-2 r}{2}} dx \ln
  \left(\frac{r+x}{(R/2)+x} \times \frac{1-R+r+x}{1-(R/2)+x}
  \right)\right]~,~  R-2 r > 0 
\label{e11} \\ 
1~,~  R-2 r =0 
\label{e22}
\end{numcases}
Thus although each term in (\ref{mult}) can be calculated (unlike for 
other block fitness distributions discussed above), most of the terms 
contribute to the distribution. As shown in 
Fig.~\ref{expoDlarge}, the distribution
${\cal G}_{s_d}$'s are quite well separated for small $d$ but get
clustered at larger $d$ which is  consistent with the behavior of
$W_d/W_{d_u}$ discussed above.

%=============================================================================
\section{Conclusions}
\label{con}

A set of random variables obtained by summing a fixed
number of i.i.d. random variables are correlated when they have at 
least one common term  in the sum. Such nonindependent variables may
describe diverse quantities such as breeding value of an animal
\cite{Rawlings:1976,Hill:1976}, fitness of a protein or antibody 
\cite{Perelson:1995} and energy of a directed polymer
\cite{Dean:2001}. For two different linear combinations of
i.i.d. random variables, we showed that as
the sum is over a set of independently distributed variables, it is 
possible to write the extreme value distribution for weakly correlated
random variables as an integral
involving the maximum value distribution of i.i.d. random
variables which makes the problem amenable to analysis at least in some
cases. Interestingly even with weak  
correlations, a rich variety of extreme value distributions result:
they can be highly 
nonuniversal parent distributions or universal extreme value
distributions for i.i.d. random variables or novel distributions unrelated to these. 

When a single mutation occurs in the initial sequence, the
limiting extreme value distribution  
(up to a rescaling) is found to be one of the i.i.d. distributions when the
initial distribution $p(f)$ decays faster than an exponential and 
parent
distribution if $p(f)$ decays slower than an exponential but a novel
distribution appears for exponentially decaying $p(f)$.  
Such a classification has also been observed in the context of
near-extreme value statistics of i.i.d. random variables 
\cite{Sabhapandit:2007b}.  
The situation is more complex when  multiple mutations are introduced in the
initial sequence. When the number of mutations $D$ does not scale with $\ell$ so that $D/\ell \to 0$ for large $\ell$, our analysis showed that $D=1$ and $1 < D  \ll \ell$ may
exhibit different distributions. While
the algebraically decaying parent distributions remain robust in that the universal Fr{\'e}chet distribution holds for $1 \leq D \ll \ell$,  for bounded distributions the
nonuniversal parent distribution for single mutation problem changed to universal Weibull distribution  for $D >1$.  For unbounded, non-exponential parent distributions decaying faster than a
power law also, the extreme value distribution for single mutation case fail to hold for $D> 1$. For all the three classes of block fitness distributions considered, novel extreme value distributions are expected if the number of mutations in both blocks is comparable to the block length. For exponentially distributed block fitnesses, we analysed the maximum value distribution for all $D$ and gave explicit
expressions for two new 
extreme value distributions (see (\ref{D1exact}) and 
(\ref{err})).  It would be very interesting to see if they occur in
other extreme value problems as well. 

{\bf Acknowledgments:}
The author thanks S. Sabhapandit for pointing out reference \cite{Sabhapandit:2007b}.   \\

%=============================================================================
%BIBLIOGRAPHY
%=============================================================================

%\input{paper.bbl}
%=============================================================================
\end{document}